# Principles for data analysis workflows

Sara Stoudt[1,2,¶], Váleri N. Vásquez[1,3,¶], Ciera C. Martinez[1,4,¶*]

[1] Berkeley Institute for Data Science, University of California Berkeley, Berkeley, CA, USA
[2] Statistical & Data Sciences Program, Smith College, Northampton, MA, USA
[3] Energy and Resources Group, University of California Berkeley, Berkeley, CA, USA
[4] Department of Molecular and Cellular Biology, University of California Berkeley, Berkeley, CA, USA

*Corresponding author
email: ccmartinez@berkeley.edu

[¶] All authors contributed equally

## Abstract

Traditional data science education often omits training on research workflows: the process that moves a scientific investigation from raw data to coherent research question to insightful contribution. In this paper, we elaborate basic principles of a reproducible data analysis workflow by defining three phases: the Exploratory, Refinement, and Polishing Phases. Each workflow phase is roughly centered around the audience to whom research decisions, methodologies, and results are being immediately communicated. Importantly, each phase can also give rise to a number of research products beyond traditional academic publications. Where relevant, we draw analogies between principles for data-intensive research workflows and established practice in software development**.** The guidance provided here is not intended to be a strict rulebook; rather, the suggestions for practices and tools to advance reproducible, sound data-intensive analysis may furnish support for both students and current professionals.

# Introduction

Traditional data science education includes a review of various statistical analysis methods as well as training in **computational tool**s, software, and programming languages. However, the development and pursuit of a research **workflow** -- the process that moves a scientific investigation from raw data to coherent question to insightful contribution -- is a crucial component of pragmatic data science that is often left out of classroom discussions. Too frequently, students and practitioners of data science are left to learn these essential skills on their own and on the job. Guidance on the breadth of potential products that can emerge from research is also lacking. In the interest of both **reproducible** science and effective career-building, data science instruction must prime researchers to regularly generate outputs over the course of their workflow.

High-quality, comprehensive education and training in **data-intensive research** should include guidance on workflow; specifically, on the creation and standardization of practices to organize data and code such that they are reproducible and culminate in results that both constitute a scientific contribution and are able to be communicated. To be high-impact, or even useful, research analyses must be contextualized in the data processing decisions that led to their creation and accompanied by a narrative that explains why the rest of the world should be interested. One way of thinking about this is that the scientific method must be tangibly reflected, and feasibly reproducible, in any data-intensive research project.

Discussions of workflow in data science often get conflated with **pipeline** development in software engineering, which is described as a series of processes that can be programmatically defined and automated. Pipelines are usually explained in the context of inputs and outputs. However, there is an important distinction between pipelines and workflows: the former refers to what a computer does, for example when a piece of software automatically runs a series of Bash or R **script**s. Meanwhile, a workflow describes what a researcher does to make advances on scientific questions: developing hypotheses, wrangling data, writing code, and interpreting results. Here, "data-intensive" research is used interchangeably with "data science" in a recognition of the breadth of domain applications that draw upon computational analysis methods and workflows.

Data analysis workflows can culminate in a number of outcomes that are not restricted to the traditional products of software engineering (software tools and packages) or academia (research papers). Rather, the workflow that a researcher defines and iterates over the course of a data science project can lead to intellectual contributions as varied as novel data sets, new methodological approaches, or teaching materials in addition to the classical tools, packages, and papers. While the workflow should be designed to serve the researcher and their collaborators, maintaining a structured approach throughout the process will inform results that are **replicable** and easily translated into a variety of products that furnish scientific insights for broader consumption.

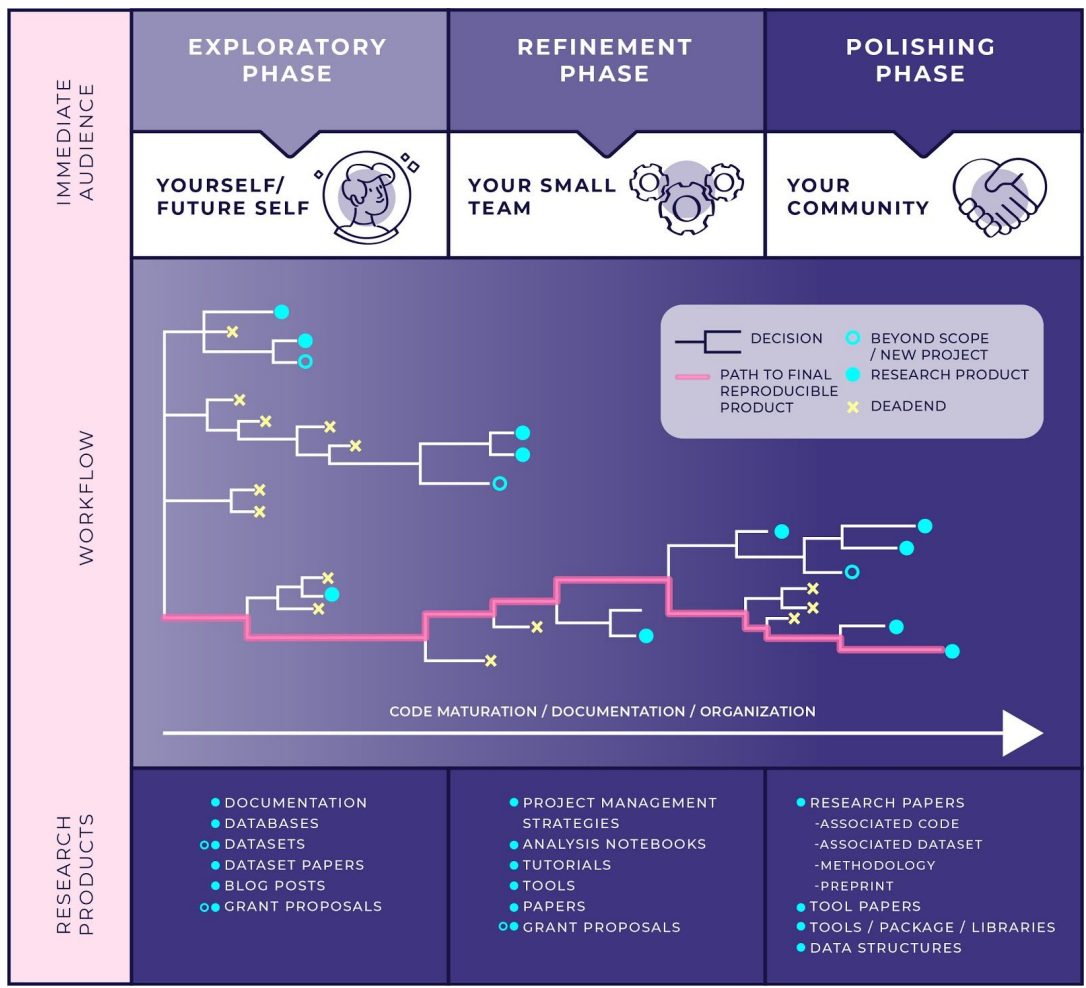

*Figure 1: Schematic of a conceptual data analysis workflow. We describe three phases of a data analysis project, each largely defined by the immediate audience to whom project decisions, methodology, and results are being communicated. Each phase can result in research products (turquoise circles) for your community in the Polishing phase, that provide knowledge and can advance career goals. Each branch in the tree represents a decision that needs to be made about the project, such as the removal of data, the use of a particular tool or model, etc. Throughout the natural life of a project there are many dead ends. These may include choices that do not work, such as experimentation with a tool that is ultimately not compatible with our data (yellow Xs). However, there are also dead ends that lie beyond the scope of our current project but may turn into a new project later on (open turquoise circles). Importantly, researchers may iterate many times between the three workflow phases during a given data analysis project.*

In the following sections, we explain the basic principles of a constructive and productive data analysis workflow by defining three phases: the Exploratory, Refinement, and Polishing Phases. Each phase is roughly centered around the audience to whom research decisions, methodologies, and results are being immediately communicated. Where relevant, we draw analogies to the realm of **software development**. While the three phases described here are not intended to be a strict rulebook, we hope that the many references to additional resources -- and suggestions for non-traditional research products -- provide guidance and support for both students *and* current data science practitioners.

## A Data Analysis Workflow in Three Phases

We partition the workflow of a data-intensive research process into three phases: Exploratory, Refinement, and Polishing. These phases are visually described in **Figure 1**. In the Exploratory Phase, researchers "meet" their data: process it, interrogate it, and sift through potential solutions to a problem of interest. In the Refinement Phase, researchers narrow their focus to a particularly promising approach, develop prototypes, and organize their code into a clearer narrative. In the Polishing Phase, researchers prepare their work for broader consumption and critique.

Each phase has an immediate audience -- the researcher themselves, their collaborative groups, or the public -- that broadens progressively and guides priorities. Each of the three phases can also benefit from principles that the software development community uses to streamline their code-based pipelines; many such standards and practices can be adapted to help structure a data-intensive researcher's workflow. Each phase also has the potential to produce a variety of research products, a prospect that we hope motivates researchers to impose these principles for data analysis from the outset of any project.

## Phase 1: Exploratory

Data-intensive research projects typically start with a domain-specific question or a particular dataset to explore (1).  The often messy Exploratory Phase is rarely discussed as an explicit step of the methodological process, but it is an essential component of research: it allows us to gain intuition about our data, informing future phases of the workflow. As we explore our data, we refine our research question and work towards the articulation of a well-defined problem. The following section will address how to reap the benefits of dataset and problem space exploration, and provide pointers on how to impose structure and reproducibility during this inherently creative phase of the research workflow.

**Designing Data Analysis: Goals and Standards of the Exploratory Phase**

Trial and error is the hallmark of the Exploratory Phase. In  "Designerly Ways of Knowing" (2), the design process is described as a "co-evolution of solution and problem spaces." Like designers, data-intensive researchers explore the problem space, learn about the potential structure of the solution space, and iterate between the two spaces. Importantly, the difficulties we encounter in this phase help us build empathy for an eventual audience beyond ourselves. It is here that we experience first hand the challenges of processing our dataset, framing domain research questions appropriate to it, and structuring the beginnings of a workflow. Documenting our trial-and-error helps our own work stay on track in addition to assisting future researchers with these same issues.

One end goal of the Exploratory Phase is to determine whether new questions of interest might be answered by leveraging existing software tools (either off the shelf or with minor adjustments), versus building new computational capabilities ourselves. For example, during this phase, a common activity includes surveying the software available for our dataset or problem space and estimating its utility for the unique demands of our current analysis. Through exploration, we learn about relevant computational and analysis tools while concurrently building an understanding of our data.

A second important goal of the Exploratory Phase is cleanup, a dynamic process that often goes hand in hand with improving our understanding of the data. Once we have established the software tools -- the programming language, data analysis packages, and a handful of the useful **function**s therein -- that are best suited to our data and domain area, we can start putting those tools to use (3). We identify important variables, remove redundancies, take note of missing information, and ponder outliers in our data set. We perform initial tests, build a simple model, or create some basic visualizations to better grasp its contents and check for expected outputs. Our research is underway in earnest now, and this effort will help us to identify what questions we might be able to ask of our data.

The Exploratory Phase is often a solo endeavor. This can make navigating it difficult, especially for new researchers. It also complicates a third goal of this phase: documentation. In this phase we ourselves are our only audience, and if we are not conscientious documenters, we can easily end up concluding the phase without the ability to coherently describe our research process to that point. Record keeping in the Exploratory Phase is often subject to our individual style of approaching problems. Some styles work in "real time," subsetting or reconfiguring data as ideas occur. More methodical styles tend to systematically plan exploratory steps, recording them before taking action. These natural tendencies impact the state of our analysis code, affecting its readability and reproducibility.

However, there are strategies -- inspired by analogous software development principles -- that can help set us up for success in meeting the standards of reproducibility (4) relevant to a scientifically sound research workflow. These strategies impose a semblance of order on the Exploratory Phase. To avoid concerns of **premature optimization** (5) while we are iterating during this phase, documentation is the primary goal, rather than fine-tuning the code structure and style.

**Analogies to Software Development in the Exploratory Phase**

*Documentation: Code and Process*

Software engineers typically value formal documentation that is readable by software users. While the audience for our data analysis code may not be defined as a "software user" per se, documentation is still vital for workflow development. Documentation for data analysis workflows can come in many forms, including comments describing individual lines of code, README files orienting a reader within a code repository, descriptive commit history logs tracking the progress of code development, **docstring**s detailing function capabilities, and vignettes providing example applications. Documentation provides both a user manual for particular tools within a project (for example, data cleaning functions), and a reference log describing scientific research decisions and their rationale (for example, the reasons behind specific parameter choices).

In the Exploratory Phase, we may identify with the type of programmer described by Brant et al. as "opportunistic" (6). This type of programmer finds it challenging to prioritize documenting and organizing code that they see as impermanent or a work in progress. "Opportunistic" programmers tend to build code using others' tools, focusing on writing "glue" code that links pre-existing components, and iterate quickly. Hartmann et al. also describe this "mashup" approach (7). Rather than "opportunistic programmers," their study focuses on "opportunistic designers." This style of design "search[es] for bridges," finding connections between what first appears to be different fields. Data-intensive researchers can also use existing tools to answer our questions of interest; we tend to build our own only when needed.

Even if the code that is used for data exploration is not developed into a software-based final research product, the exploratory process as a whole should exist as a permanent record. Documenting choices and decisions we make along the way is crucial to making sure we do not forget any aspect of the analysis workflow, because each choice may ultimately impact the final results. For example, if we remove some data points from our analyses, we must know which data points we removed -- and our reason for removing them -- when we start sharing our work with others. This is an important argument against ephemerally conducting our data analysis work via the command line.

Instead of the command line, tools like a computational **notebook** (8) can help capture a researcher's decision making process in real time (9). A computational notebook where we never delete code, and -- to avoid overwriting named variables -- only move *forward* in our document, could act as "version control designed for a 10-minute scale" that Brant et al. found might help the "opportunistic" programmer. More recent advances in this area include the reactive notebook (10)(11). Such tools assist documentation while potentially enhancing our creativity during the Exploratory Phase. The bare minimum documentation of our Exploratory Phase might therefore include such a notebook or an annotated script (12) to record all analyses that we perform and code that we write.

To go a step beyond annotated scripts or notebooks, researchers might employ a **version control** system such as Git. With its issues, branches, and informative commit messages, Git is another useful way to maintain a record of our trial-and-error process and track which files are progressing towards which goals of the overall project. Using Git together with a public online hosting service such as GitHub allows us to share our work with collaborators and the public in real time, if we so choose.

A researcher dedicated to conducting an even more thoroughly documented Exploratory Phase may take Ford's advice and include notes that explicitly document our stream of consciousness (13). Our notes should be able to efficiently convey what failed, what worked but was uninteresting or beyond scope of the project, and what paths of inquiry we will continue forward with in more depth (**Figure 1**). In this way, as we transition from the Exploratory Phase to the Refinement Phase, we will have some signposts to guide our way.

*Testing: Comparing Expectations to Output*

As Ford (13) explains, we face competing goals in the Exploratory Phase: we want to get results quickly, but we also want to be confident in our answers. Her strategy is to focus on documentation over tests for "one-off" analyses that will not form part of a larger research project. However, the complete absence of formal tests may raise a red flag for some data scientists used to the concept of **test-driven development**. This is a tension between the code-based work conducted in scientific research versus software development: tests help build confidence in analysis code and convince users that it is reliable or accurate, but tests also imply finality and take time to write that we may not be willing to allocate in the experimental Exploratory Phase. However, software development-style tests do have useful analogues in data analysis efforts: we can think of tests, in the data analysis sense, as a way of checking whether our expectations match the reality of a piece of code's output.

Imagine we are looking at a dataset for the first time. What weird things can happen? The type of variable might not be what we expect (for example, the integer 4 instead of the float 4.0). The dataset could also include unexpected aspects (for example, dates formatted as strings instead of numbers). The amount of missing data may be larger than we thought, and this missingness could be coded in a variety of ways (for example, as a NaN, NULL, or -999). Finally, the dimensions of a data frame after merging or subsetting it for data cleaning may not match our expectations. These types of gaps in expectation vs. reality are "silent faults" (Hook and Kelly 2009). Without checking for them explicitly, we might proceed with analysis without noticing that anything is amiss and encode that error in our results.

For these reasons, every data exploration should include quantitative and qualitative "gut-checks" (14) that can help us diagnose an expectation mismatch as we go about examining and manipulating our data. We may check assumptions about data quality such as the proportion of missing values, verify that a joined dataset has the expected dimensions, or ascertain the statistical distributions of well-known data categories. In this latter case, having domain knowledge can help us understand what to expect. We may want to compare two datasets (for example, pre- and post-processed versions) to ensure they are the same (15); we may also evaluate diagnostic plots to assess a model's goodness of fit. Each of the elements that gut-checks help us monitor will impact the accuracy and direction of our future analyses.

We perform these manual checks to reassure ourselves that our actions at each step of data cleaning, processing, or preliminary analysis worked as expected. However, these types of checks often rely on us as researchers visually assessing output and deciding if we agree with it. As we transition to needing to convince users beyond ourselves of the correctness of our work, we may consider employing **defensive programming** techniques that help guard against specific mistakes. An example of defensive programming in the Julia language is the use of the @assert macro to validate values or function outputs. Another option includes writing "chatty functions" (16) that signal a user to pause, examine the output, and decide if they agree with it.

**When to Transition from the Exploratory Phase: Balancing Breadth and Depth**

A researcher in the Exploratory Phase experiments with a variety of potential data configurations, analysis tools, and research directions. Not all of these may bear fruit in the form of novel questions or promising preliminary findings. Learning how to find a balance between the breadth and depth of data exploration helps us understand when to transition to the Refinement Phase of data-intensive research. Specific questions to ask yourself as you prepare to transition between the Exploratory Phase and the Refinement Phase can be found in Box 2.

Imposing structure at certain points throughout the Exploratory Phase can help to balance our wide search for solutions with our deep dives into particular options. In an analogy to the software development world, we can treat our exploratory code as a code release -- the marker of a stable version of a piece of software. For example, we can take stock of the code we have written at set intervals, decide what aspects of the analysis conducted using it seem most promising, and focus our attention on more formally tuning those parts of the code. At this point, we can also note the presence of research "dead ends" and perhaps record where they fit into our thought process. As we make decisions about what research directions we are going to pursue, we can also adjust our file structure and organize files into directories with more informative names.

Just as Cross (2) finds that a "reasonably-structured process" leads to design success where "rigid, over-structured approaches" find less success, a balance between the formality of documentation and testing and the informality of creative discovery is key to the Exploratory Phase of data-intensive research. By taking inspiration from software development and adapting the principles of that arena to fit data analysis work, we add enough structure to this phase to ease transition into the next phase of the research workflow.

# Phase 2: Refinement

Inevitably, we reach a point in the Exploratory Phase when we have acquainted ourselves with our dataset, processed and cleaned it, identified interesting research questions that might be asked using it, and found the analysis tools that we prefer to apply. Having reached this important juncture, we may also wish to expand our audience from ourselves to a team of research collaborators. It is at this point that we are ready to transition to the Refinement Phase. However, we must keep in mind that new insights may bring us back to the Exploratory Phase: over the lifetime of a given research project, we are likely to cycle through each workflow phase multiple times.

In the Refinement Phase, the extension of our target audience demands a higher standard for communicating our research decisions as well as a more formal approach to organizing our workflow and documenting and testing our code. In this section, we will discuss principles for structuring our data analysis in the Refinement Phase. This phase will ultimately prepare our work for polishing into final research products, including "traditional" peer reviewed academic papers as well as a diversity of other outputs.

**Designing Data Analysis: Goals and Standards of the Refinement Phase**

The Refinement Phase encompasses many critical aspects of a research project. Additional data cleaning may be conducted, analysis methodologies are chosen, and the final experimental design is decided upon. Experimental design may include identifying case studies for variables of interest within our data. If applicable, it is during this phase that we determine the details of simulations. Preliminary results from the Exploratory Phase inform how we might improve upon or scale up prototypes in the Refinement Phase. Data management also continues to be essential; during this phase, data management may expand to include the **serialization** of experimental setups. Finally, standards of reproducibility must be maintained throughout. Each of these aspects constitutes an important goal of the Refinement Phase as we determine the most promising avenues for focusing our research workflow en route to producing polished research products that demand even higher reproducibility standards during the Polishing Phase.

All of these goals are developed in conjunction with our research team, so decisions have to be documented and communicated in a way that is reproducible and constructive within that group. Just as the solitary nature of the Exploratory Phase can be daunting, the collaboration that happens in the Refinement Phase brings its own set of challenges as we figure out how to best work together. Our team can be defined as the people who participate in developing the research question, preparing the dataset it is applied to, coding the analysis, or interpreting results. It might also include individuals who offer feedback about the progress of our work. In the context of academia, our team usually includes our laboratory or research group. Like most other aspects of data-intensive research, our team may evolve as the project evolves. But however we define our team, its members inform how our efforts proceed during the Refinement Phase: thus, another primary goal of the Refinement Phase is establishing group-based standards for the research workflow. Specific questions to ask yourself during this phase can be found in Box 2.

In recent years, the conversation on standards within academic data science and scientific computing has shifted from "best" practices (17) to "good enough" practices (18). This is an important distinction when establishing team standards during the Refinement Phase: reproducibility is a spectrum (19), and collaborative work in data-intensive research carries unique demands on researchers as scholars and coworkers (20). At this point in the research workflow, standards should be adopted according to their appropriateness for our team. This means talking amongst ourselves not only about scientific results, but also about the experimental design that led to those results and the role that each team member plays in the research workflow. Establishing methods for effective communication is therefore another important goal in the Refinement Phase, as we cannot develop group-based standards for the research workflow without it.

**Analogies to Software Development in the Refinement Phase**

*Documentation as a Driver of Reproducibility*

The concept of literate programming (Knuth 1984) is at the core of an effective Refinement Phase. This philosophy brings together code with human-readable explanations, allowing scientists to demonstrate the functionality of their code in the context of words and visualizations that describe the rationale for and results of their analysis. Computational notebooks are useful in the Exploratory Phase and are also applicable here, where they can assist with team-wide discussions, research development, prototyping, and idea sharing. Jupyter Notebooks (21) are agnostic to choice of programming language and so provide a good option for research teams that may be working with a diverse code base or different levels of comfort with a particular programming language. Language-specific interfaces such as R's RMarkdown functionality (22) or the reactive notebook put forward by Pluto.jl in the Julia programming language furnish additional options for literate programming.

The same strategies that promote scientific reproducibility for traditional laboratory notebooks can be applied to the computational notebook (23). After all, our data-intensive research workflow can be considered a sort of scientific experiment -- we develop a hypothesis, query data, support or reject our hypothesis, and state our insights. A central tenet of scientific reproducibility is recording inputs relevant to a given analysis, such as parameter choices, and explaining any calculation used to obtain them, so that our outputs can later be verifiably replicated. Methodological details -- for example the decision to develop a dynamic model in continuous time versus discrete time, or the choice of a specific statistical analysis over alternative options -- should also be fully explained in computational notebooks developed during the Refinement Phase. Domain knowledge may inform such decisions, making this an important part of proper notebook documentation; such details should also be elaborated in the final research product. Computational research descriptions in academic journals generally include a narrative relevant to their final results, but these descriptions often do not include enough methodological detail to enable replicability, much less reproducibility. However, this is changing with time (24,25).

As scientists, we must keep a record of the tools we use to obtain our results in addition to our methodological process. In a data-intensive research workflow, this includes documenting the specific version of any software that we used, as well as its relevant dependencies and compatibility constraints. Recording this information at the top of the computational notebook that details our data science experiment allows future researchers -- including ourselves and our teams -- to establish the precise computational environment that was used to run the original research analysis. Our chosen programming language may supply automated approaches for doing this, such as a **package manager**, simplifying matters and painlessly raising the standards of reproducibility in a research team. The unprecedented levels of reproducibility possible in modern computational environments have produced some variance in the expectations of different research communities; it behooves the research team to investigate the community-level standards applicable to our specific domain science and chosen programming language.

A notebook can include more than a deep dive into a full-fledged data science experiment. It can also involve exploring and communicating basic properties of the data, whether for purposes of training team members new to the project or for brainstorming alternative possible approaches to a piece of research. In the Exploration Phase we have discovered aspects that we want our research team to know about, for example outliers or unexpected distributions, and created preliminary visualizations to better understand their presence. In the Refinement Phase, we may choose to improve these initial plots and reprise our data processing decisions with team members to ensure that the logic we applied still holds.

Computational notebooks can live in private or public repositories to ensure accessibility and transparency among team members. A version control system such as Git continues to be broadly useful for documentation purposes in the Refinement Phase, beyond acting as a storage site for computational notebooks. Especially as our team and code base grows larger, a history of commits and pull requests helps keep track of responsibilities, coding or data issues, and general workflow.

Importantly, all tools have their appropriate use cases. Researchers must not develop an over reliance on computational notebooks and should learn to recognize when different tools are required. Computational notebooks may quickly become unwieldy for certain projects and large teams, incurring **technical debt** in the form of duplications or over-written variables. As code grows in complexity over the course of a data analysis, keeping track of changes becomes more error prone. Computational notebooks are also memory intensive, and this usage increases at a rate unsustainable for large and sometimes even medium-sized data projects. As our research project grows in complexity and size, or gains team members, transitioning to an **Integrated Development Environment** (IDE) such as Pycharm or a **source code editor** such as Visual Studio Code -- which interact easily with **container** environments like Docker and version control systems such as GitHub -- can help scale our data analysis, while still retaining important properties like reproducibility.

*Testing and Establishing Code Modularity*

Code in data-intensive research is generally written as a means to an end, the end being a scientific result from which researchers can draw conclusions. This stands in stark contrast to the purpose of code developed by data engineers or computer scientists, which is generally written to optimize a mechanistic function for maximum efficiency. During the Refinement Phase, we may find ourselves with both analysis-relevant *and* **mechanistic code**, especially in "big data" statistical analyses or complex dynamic simulations where optimized computation becomes a concern. Keeping the immediate audience of this workflow phase, our research team, at the forefront of our mind can help us take steps to structure both mechanistic and analysis code in a useful way.

Mechanistic code, which is designed for repeated use, often employs abstractions by wrapping code into functions that apply the same action repeatedly or stringing together multiple scripts into a computational pipeline. **Unit test**s and so-called **accessor function**s or **getter and setter function**s that extract parameter values from **data structures** or set new values are examples of mechanistic code that might be included in a data-intensive research analysis. Meanwhile, code that is designed to gain statistical insight into distributions or model scientific dynamics using mathematical equations are two examples of analysis code. Sometimes, the line between mechanistic code and analysis code can be a blurry one. For example, we might write a looping function to sample our dataset repeatedly, and that would classify as mechanistic code. But that sampling may be designed to occur according to an algorithm such as Markov Chain Monte Carlo (MCMC) that is directly tied to our desire to sample from a specific probability distribution; therefore, this could be labeled analysis code.

It is common practice to wrap code that we use repeatedly into functions to increase readability and **modularity** while reducing the propensity for user-induced error. However, the scripts and programming notebooks so useful to establishing a narrative and documenting work in the Refinement Phase are set up to be read in a linear fashion. Embedding mechanistic functions in the midst of the research narrative obscures the utility of the notebooks in telling the research story, and generally clutters up the analysis with a lot of extra code. For example, if we develop a function to eliminate the redundancy of repeatedly restructuring our data to produce a particular type of plot, we do not need to showcase that function in the middle of a computational notebook analyzing the implications of the plot that is created -- the point is the research implications of the image, not the code that made the plot. Then where do we keep the data-reshaping, plot-generating code?

Strategies to structure the more mechanistic aspects of our analysis can be drawn from common software development practices. As our team grows or changes, we may require the same mechanistic code. For example, the same data-reshaping, plot-generating function described earlier could need to be pulled into multiple computational experiments that are set up in different locations, computational notebooks, scripts, or Git branches. Therefore, a useful approach would be to start collecting those mechanistic functions into their own script or file, sometimes called "helpers" or "utils", that acts as a supplement to the various ongoing experiments, wherever they may be conducted. This separate script or file can be referenced or "called" at the beginning of the individual data analyses. Doing so allows team members to benefit from collaborative improvements to the mechanistic code without having to re-invent the wheel themselves. It also preserves the narrative properties of team members' analysis-centric computational notebooks or scripts while maintaining transparency in basic methodologies that ensure project-wide reproducibility. The need to begin collecting mechanistic functions into files separate from analysis code is a good indicator that it may be time for the research team to transition away from computational notebooks and towards a code editor or IDE.

Testing scientific software is not always perfectly analogous to testing typical software development projects, where automated **continuous integration** is often sufficient (26). However, as we start to modularize our code, breaking it into functions and from there into separate scripts or files that serve specific purposes, principles from software engineering become more readily applicable to our data-intensive analysis. Unit tests can now help us ensure that our mechanistic functions are working as expected, formalizing the "**gut check**s" we performed in the Exploratory Phase. Among other applications, these tests should verify that our functions return the appropriate value, object type, or error message as needed (27). Formal tests can also provide a more extensive investigation of how "trustworthy" the performance of a particular analysis method might be, affording us an opportunity to check the correctness of our scientific inferences.

**When to Transition from the Refinement Phase: Going Backwards and Forwards**

Workflows in data science are rarely linear; it is often necessary for researchers to iterate between the Refinement and Exploratory Phases. For example, while our research team may have decided on an experimental design to pursue in the Refinement Phase, the scope of that design may require us to revisit decisions made during the data processing that was conducted in the Exploratory Phase. This might mean including additional information from supplementary datasets to help refine our hypothesis or research question. In returning to the Exploratory Phase, we investigate these potential new datasets and decide if it makes sense to merge them with our original dataset.

Iteration between the Refinement and Exploratory Phases is a careful balance. On the one hand, we must be careful not to allow "scope creep" to expand our problem space beyond an area where we are able to develop constructive research contributions. On the other hand, if we are too rigid about decisions made over the course of our workflow and refuse to look backwards as well as forwards, we may risk cutting ourselves off from an important part of the potential solution space.

Data-intensive researchers can once more look to principles within the software development community to help guide the careful balancing act required to conduct research that is both comprehensive and able to be completed: teams can develop data or project management plans to define desirable goals and deliverable dates. Because these plans can serve as research products themselves, they are described in further detail in the next phase of the workflow -- Polishing.

## Phase 3: Polishing

In the previous sections of this paper, we discussed how to progress from the exploration of raw data through the refinement of a research question and selection of an analytical methodology. We also described how the details of that workflow are guided by the breadth of the immediately relevant audience: ourselves in the Exploratory Phase, and our research team in the Refinement Phase. In the Polishing Phase, it becomes time to make our data analysis camera-ready for a much broader group. This may translate to developing a variety of research products in addition to -- or instead of -- traditional academic outputs like peer reviewed publications and typical software development products such as computational tools.

**Beyond Data Analysis: Goals and Standards of the Polishing Phase**

The main goal of the Polishing Phase is to prepare our analysis to enter the public realm as a set of products ready for external use, reflection, and improvement. "Polishing" encompasses the cleanup that happens prior to initially sharing our results (for example, ahead of submitting our work to peer review). It also includes the process of incorporating suggestions for improvement prior to finalization (for example, adjustments to address reviewer comments ahead of publication). The research products that emerge from a given workflow may vary in both their form and their formality -- indeed, some research products, like a code base, might continually evolve without ever assuming "final" status -- but each product constitutes important contributions that push scientific boundaries in their own way.

Importantly, producing polished outcomes over the course of an entire workflow rather than just at the end of a project can help researchers progressively build their data analysis portfolios and fulfill a second goal of the Polishing Phase: gaining credit, and credibility, in our domain area. This is especially relevant for junior scientists who are just starting research careers or wish to become industry data scientists (1). Developing polished products at several intervals along a single workflow is also instructional for the researcher themselves. Researchers who prepare their work for public assessment from the earliest phases of an analysis become acquainted with the pertinent problem and solution spaces from multiple perspectives. This additional understanding, together with the feedback that polished products generate from people outside our immediate team, may furnish insights that improve our approach in other phases of the research workflow.

Building our data science research portfolio requires a method for tracking and attributing the many products that we might develop. One important method for tracking and attribution is the digital object identifier, or DOI. It is a unique handle, standardized by the International Organization for Standardization (ISO), that can be assigned to different types of information objects. DOIs are usually connected to metadata, for example, they might include a URL pointing to where the object they are associated with can be found online. Academic researchers are used to thinking of DOIs as persistent identifiers for peer reviewed publications. However, DOIs can also be generated for datasets, GitHub repositories, computational notebooks, teaching materials, management plans, reports, **white papers**, and pre-prints.

A third, longer-term goal of the Polishing Phase involves establishing a researcher's professional trajectory. Every individual needs to gauge how their compendium of research products contribute to their career, and how intentional portfolio-building might in turn drive the research that they ultimately conduct. For example, researchers who wish to work in academia might feel obliged to obtain "academic value" from less traditional research products by essentially reprising them as peer reviewed papers. But judging a researcher's productivity by the metric of paper authorship can alter how and even whether research is performed (28). Increasingly, academic journals are revisiting their publishing requirements (29) and raising standards of reproducibility. This fact is bringing the data and programming methodologies that underpin written analyses closer to center stage. Data-intensive research, and the people who produce it, stand to benefit. Scientists -- now encouraged, and even required by some academic journals to share data and code -- can publish and receive credit as well as feedback for the multiple research products that support their publications. Questions to ask ourselves as we consider possible products can be found in Box 2.

**Polishing: Products of the Exploratory Phase**

The old adage that one person's trash is another's treasure is relevant to the Exploratory Phase of a data science analysis: of the many potential applications for a particular data set, there is often only time to explore a small subset. Those applications which fall outside the scope of the current analysis can nonetheless be valuable to our future selves, or to others seeking to conduct their own analyses. To that end, the documentation that accompanies data exploration can furnish valuable guidance for later projects. Further, the cleaned and processed data set that emerges from the Exploratory Phase is itself a valuable outcome that can be assigned a DOI and published as a formal product of this portion of the data analysis workflow, using outlets like Dryad (http://www.**datadryad**.org), Figshare (https://figshare.com/), Open Science Framework (https://osf.io/), Zenodo (https://zenodo.org/), or re3data (https://www.re3data.org/) among others. Publishing the dataset, along with its metadata, is an essential component of scientific transparency and reproducibility; it is fundamentally valuable to the scientific community.

The Git repositories or computational notebooks that archive a data scientist's approach, uncover coding bugs, redundancies, or inconsistencies, and note the rationale for focusing on specific aspects of the data are also useful research products in their own right. These items can provide a touchstone for alternative explorations of the same dataset at a later time. In addition to documenting valuable lessons learned, contributions of this kind can formally augment a data scientist's registered body of work: code used to actively clean data or record an exploratory process can be made citable by employing services like Zenodo to add a DOI to the applicable Git repository. Smaller code snippets or data excerpts can be shared – publicly or privately – using the more lightweight GitHub Gists ([https://gist.github.com/](https://gist.github.com/)). Tools such as Dr.Watson ([https://github.com/JuliaDynamics/DrWatson.jl](https://github.com/JuliaDynamics/DrWatson.jl)), which are designed to assist researchers with organization and reproducibility and even include workflow tutorials for scientific data projects, can inform the polishing process for products emerging from any phase of the analysis. Additional tools and platforms in this vein include Snakemake (https://snakemake.readthedocs.io/), Nextflow (https://www.nextflow.io), Common workflow language (https://www.commonwl.org), and Workflow description language (https://openwdl.org)

Alternative mechanisms for crediting the time and talent that researchers invest in the Exploratory Phase include relatively informal products. For example, blog posts can detail problem space exploration for a specific research question or lessons learned about data analysis training and techniques. White papers that describe the raw dataset and the steps taken to clean it, together with an explanation of why and how these decisions were taken, might constitute another such informal product. Versions of these blog posts or white papers can be uploaded to open-access websites such as [arXiv.org](arXiv.org) as preprints, and receive a DOI.

The familiar academic route of a peer-reviewed publication is also available for products emerging from the Exploratory Phase. For example, depending on the domain area of interest, journals such as Nature Scientific Data and IEEE Transactions are especially suited to papers that document the methods of dataset development or simply reproduce the dataset itself. Pedagogical contributions learned or applied over the course of a research workflow can be written for submission to training-focused journals such as the Journal of Statistics Education. For a list of potential product examples that can come out of this phase, see Box 3.

**Polishing: Products of the Refinement Phase**

In the Refinement Phase, documentation and the ability to communicate both methods and results become essential to daily management of the project. Happily, the implementation of these basic practices can also provide benefits beyond the immediate team of research collaborators: they can be standardized as a Data Management Plan or Protocol (DMP). DMPs are a valuable product that can emerge from the Refinement phase as a formal version of "lessons learned" concerning both research and team management. This product records the strategies and approaches used to, for example, enact data description, sharing, storage, analysis, and preservation.

While DMPs are often living documents over the course of a research project, evolving dynamically with the needs or restrictions that are encountered along the way, there is great utility to codifying them either for our team's later use or for others conducting similar projects. DMPs can also potentially be leveraged into new research grants for our team, as these protocols are now a common mandate by many funders (30). The group discussions that contribute to developing a DMP can be difficult, and encompass considerations relevant to everything from team building to research design. The outcome of these discussions are often directly tied to the constructiveness of a research team and its robustness to potential turnover (30). Sharing these standards and lessons learned in the form of polished research products can propel a proactive discussion of data management and sharing practices within our research domain. This in turn bolsters the creation or enhancement of community standards beyond our team, and provides training materials for those new to the field.

As with the research products that are generated by the Exploratory Phase, DMPs can lead to polished blog posts, training materials, white papers, and pre-prints that enable researchers to both spread the word about their valuable findings and be credited for their work. In addition, peer reviewed journals are beginning to allow the publication of DMPs as a formal outcome of the data analysis workflow (e.g. Rio Journal). Importantly, when new members join a research team, they should receive a copy of the group's DMP. If any additional training pertinent to plans or protocols is furnished to help get new members up to speed, these materials too can be polished into research products that contribute to scientific advancement. For a list of potential product examples that can come out of this phase, see the Box 3.

**Polishing: Traditional Research Products**

By polishing our work, we finalize and format it to receive critiques beyond ourselves and our immediate team. The scientific analysis and results that are born of the full research workflow -- once documented and linked appropriately to the code and data used to conduct it -- are most frequently packaged into the traditional academic research product: peer-reviewed publications. Even this product, however, can be improved upon vis-à-vis its reproducibility and transparency thanks to modern tools. For example, papers that employ literate programming notebooks enable researchers to augment the real-time evolution of a written draft with the code that informs it.

Peer-reviewed papers are of primary importance to the career and reputation of academic researchers (31), but the traditional format for such publications often does not take into account essential aspects of data-intensive analysis such as computational reproducibility (32). Where strict requirements for reproducibility are not enforced by a given journal, researchers should nonetheless compile the supporting products that made our submitted manuscript possible -- including relevant code and data, as well as the documentation of our computational tools and methodologies as described in the earlier sections of this paper -- into a research compendium (29,33–35). The objective is to provide transparency to those who read our academic publication, and reproduce the workflow that led to our results **(Figure 1)**.

In addition to peer-reviewed publications and the various alternative research products described above, some scientists may choose to revisit the scripts developed during the Exploratory or Refinement Phases and polish that code into a traditional software development product: a computational tool or **software tool**. A computational tool can include libraries, packages, collections of functions, or data structures designed to help with a specific class of problem. Such products might be accompanied by repository documentation or a full-fledged methodological paper that can be categorized as additional research products beyond the tool itself. Each of these items can augment a researcher's body of citable work and contribute to advances in our domain science. One very simple example of a tool might be an interactive web application built in RShiny (https://shiny.rstudio.com/) that allows the easy exploration of cleaned data sets, or demonstrates the outcomes of alternative research questions. More complex examples include a software package that builds an open source data analysis pipeline, or a data structure that formally standardizes the problem space of a domain-specific research area.

While the software engineering literature furnishes a rich suite of resources for researchers seeking to develop their own computational tools, this existing body of work is generally directed toward trained programmers and software engineers. The design decisions that are crucial to scientists – who are primarily interested in data analysis, experiment **extensibility**, and result reporting and inference – can be obscured by concepts that are either out of scope or described in overtly technical jargon. In Box 4 we furnish a basic guide designed to highlight the decision points and architectural choices relevant to creating a tool for data-intensive research. Data scientists seeking to wade into computational tool development are well advised to review the guidelines described in Gruning et al 2019 (36) in addition to more traditional software development resources and texts such as Clean Code (37), Refactoring (38), and Best Practices in Scientific Computing (17).

## Conclusion

Defining standards for data analysis workflows is important for scientific accuracy, efficiency, and the effective communication of results, regardless of whether we are working alone or in a research team. Establishing standards assists practicing data science researchers and sets expectations to help fledgling data scientists learn the importance of computational reproducibility from the outset of their careers. There is no single set of principles for "correctly" performing data-intensive research. Each computational project carries its own context -- from the scientific domain in which it is conducted, to the software and methodological analysis tools we use to pursue our research questions, to the dynamics of our particular research team. Therefore, this paper has outlined some general principles for thinking through the design of a data analysis workflow such that researchers may consider options that might work for them. It has also put forward suggestions for research products that might emerge from each phase of a data analysis workflow.

Aiming for full reproducibility when communicating research results is a noble pursuit, but it is imperative to understand that there is a balance between generating a complete analysis and furnishing a 100% reproducible product. Researchers have competing motivations: finishing their work in a timely fashion versus having a perfectly documented final product, all while balancing how their work can be used to strengthen their career. Despite various calls for the creation of a standard framework (4,39), achieving complete reproducibility may go far beyond the researcher to encompass how analysis and version control software tracks workflow, as well as a culture-wide shift in how research workflows are documented. Both of these advancements are challenging and are unlikely to manifest quickly, although they are underway across a number of scientific domains (19). By reframing what a formal research product can be -- and noting that polished contributions can constitute much more than the academic publications previously held forth as the benchmark for career advancement -- we motivate structural change to data analysis workflows.

As researchers design data analysis workflows, it is important to keep the knowledge that research products can emerge from every phase at the forefront of our efforts. By doing so, researchers can amass outputs far beyond the academic paper. Increasingly, there are venues for writing less traditional papers that describe or consist solely of a novel dataset, a software tool, a particular methodology, or training materials . As the professional landscape for data-intensive research evolves, these less traditional research publications and products are extremely valuable for distinguishing applicants to academic *and* non-academic jobs, grants, and teaching positions. Data scientists and researchers must possess numerous and multifaceted skills to perform scientifically robust and computationally effective data analysis. Therefore, potential research collaborators or hiring entities both inside and outside the academy should take into account a variety of research products, from every phase of the data analysis workflow, when evaluating the career performance of data-intensive researchers (40).

The ever-growing landscape of data-intensive research, and its emphasis on reproducibility, corresponds in many ways to the growing importance of science communication in the broader world. Science communication can connect data-intensive research -- whether conducted in academia or elsewhere -- to policymakers, domain professionals, and laypeople (41–43). Establishing community standards and principles for data analysis workflows can help researchers stress the importance of science communication in their respective domains and communities. Such principles advocate and explain the value of our work through narrative, helping to link the seemingly esoteric methods or concepts applied in research with real-world change.

## Author Contributions

SAS, VNV, and CCM researched, wrote, and edited this work.

## Acknowledgments

We thank the Best Meta Working Group (UC Berkeley) for the thoughtful conversations and feedback that greatly informed the content of this paper. We thank the Berkeley Institute for Data Science for hosting meetings that brought together data scientists, biologists, statisticians, computer scientists, and software engineers to discuss how data-intensive research is performed and evaluated. We especially thank Stuart Gieger (UC Berkeley) for his leadership of the Best Meta Group and Rebecca Barter (UC Berkeley) for her feedback and edits on earlier drafts of the manuscript. SAS was supported by the National Physical Sciences Consortium fellowship. SAS, VNV, and CCM were supported by the Gordon & Betty Moore Foundation (GBMF3834) and Alfred P. Sloan Foundation (2013-10-27) as part of the Moore-Sloan Data Science Environments. CCM holds a Postdoctoral Enrichment Program Award from the Burroughs Wellcome Fund.

# Boxes

## Box 1: Terminology

***Box 1 caption:*** *This box provides definitions for terms in **bold** throughout the text. Terms are sorted alphabetically and cross referenced where applicable.*

**Accessor function:** A function that returns the value of a variable. Also called a "getter function." See "mutator method."

**Assertion:** An expression that is expected to be true at a particular point in the code.

**Computational tool:** May include libraries, packages, collections of functions, and/or data structures that have been consciously designed to facilitate the development and pursuit of data-intensive questions.

**Software tool:** Often used as a synonym for computational tool.

Container**:** The bundle of information and resources needed to reproduce a computational environment, including but not limited to software versions,applications, and dependencies.

**Continuous integration:** Automatic tests that update code against pre-established tests and checks.

**Gut check:** Also "data gut check." Quick, broad, and shallow testing (44) before and during data analysis. Synonymous words: smoke test, sanity check (45), consistency check, sniff test, soundness check. Although these terms are usually described in the context of software development, the concept of a data-specific gut check has not been described. Examples include checking change of dimensions data structures after merging data, assessing null values/missing values, zero values, negative values, and range of values to see if they make sense.

**Data-intensive research**: Research that is centrally based on the analysis of data and its structural or statistical properties. May include but is not limited to research that hinges on large volumes of data or a wide variety of data types; the modern era of computing has seen a rapid rise in the amount of data available to inform research questions.

**Data science research**: Often used as a synonym for data-intensive research. "Data science" as a standalone term may also refer more broadly to the use of computational tools and statistical methods to gain insights from digitized information.

**Data structure:** A format for storing data values and definition of operations that can be applied to data of a particular type.

**Defensive programming**: Strategies to guard against failures or bugs in code; this includes the use of tests and assertions.

**Docstring:** A code comment for a particular line of code.

**Extensibility:** The flexibility to be extended or repurposed in a new scenario.

**Function:** A piece of more abstracted code that can be reused to perform the same operation on different inputs of the same type and has a standardized output.

(46–48)

**Getter function:** Another commonly used term for an accessor function.

**Integrated development environment (IDE):** A software application that facilitates software development and minimally consists of a source code editor, build automation tools, and a debugger.

**Modularity:** An ability to separate different functionality into stand-alone pieces.

**Mutator method:** A function used to control changes to variables. See "setter function" and "accessor function."

**Notebook:** A computational or physical place to store details of a research process including decisions made.

**Mechanistic code**: Code used to perform a task as opposed to conduct an analysis. Examples include processing functions and plotting functions.

**Overwrite:** The process, intentional or accidental, of assigning new values to existing variables.

**Package manager:** A system used to automate the installation and configuration of software.

**Pipeline**: A series of programmatic processes during data analysis and data cleaning, usually linear in nature, that can be automated and usually be described in the context of inputs and outputs.

**Premature optimization**: Focusing on details before the general scheme is decided upon.

**Refactoring:** A change in code to make it more organized or efficient without changing the overall output.

**Replicable:** A new study arrives at the same scientific findings as a previous study, collecting new data (with the same or different methods) and completes new analyses. (49–51).

**Reproducible:** Authors provide all the necessary data and the computer codes to run the analysis again, recreating the results. (49–51).

**Script**: A collection of code, ideally related to one particular step in the data analysis.

**Setter function:** A type of function that controls changes to variables. It is used to directly access and alter specific values.

**Serialization:** The process of saving data structures, inputs and outputs, and experimental setups generally in a storable, shareable format. Serialized information can be reconstructed in different computer environments for the purpose of replicating or reproducing experiments.

**Software development:** A process of writing and documenting code in pursuit of an end goal, typically focused on process over analysis.

**Source code editor:** A program that facilitates changes to code by an author.

**Technical debt:** The extra work you defer by pursuing an easier, yet not ideal solution, early on in the coding process.

**Test-driven development:** Each change in code must be verified against tests to prove its functionality.

**Unit test:** A code test for the smallest chunk of code that is actually testable.

**Version control:** A way of managing changes to code or documentation that maintains a record of changes over time.

**White paper:** An informative, at least semi-formal document that explains a particular issue but is not peer-reviewed.

**Workflow**: The process that moves a scientific investigation from raw data to coherent research question to insightful contribution. This often involves a complex series of processes, and includes a mixture of machine automation and human intervention. It is a nonlinear and iterative exercise.

## Box 2: Questions

***Box 2 caption:*** *This box provides guiding questions to assist readers in navigating through each workflow phase. Questions pertain to planning, organization, and accountability over the course of workflow iteration.*

Questions to ask in the Exploratory Phase

- Who can read through our materials and understand our workflow?
    - Good: Ourselves (e.g. Code includes signposts refreshing our memory of what is happening where.)
    - Better: Our small team who has specialized knowledge about the context of the problem.
    - Best: Anyone with experience using similar tools to us.
- What material do we think is worth continuing into the next phase?
    - Good: Dead ends marked differently than relevant and working code.
    - Better: Material connected to a handful of promising leads.
    - Best: Material connected to a clearly defined scope.
- Where does the code live?
    - Good: Backed up in a second location in addition to our computer.
    - Better: Within a shared space amongst our team (e.g. Google Drive, Box, etc.).
    - Best: Within a version control system (e.g. GitHub) that furnishes a complete timeline of actions taken.
- Why did we make particular data cleaning and analysis decisions?
    - Good: Noted in a separate place from our code (e.g. a physical notebook).
    - Better: Noted in comments throughout the code itself, with expectations informally checked.
    - Best: Noted systematically throughout code as part of a narrative, with expectations formally checked.

Questions to ask in the Refinement Phase

- Who is in our team?
- What are our teammates' skill levels?
    - Consider career level, computational experience, and domain specific experience.
- How do we communicate methodology with our teammates' skills in mind?

- Are there established standards within the team that need to be adopted, or conflict with our Exploration Phase workflow?
    - What reproducibility tools can be agreed upon?
- What is the main takeaway of our findings?
    - How can our work be packaged into impactful research products?
    - Can we explain the same important results across different platforms (e.g. blog post in addition to white paper)?
- Who will be affected by the outcomes of our work?
    - How can we alert these people and make our work accessible?
- Why is our work important for our domain-specific field? For broader society?
    - How can we use narrative to make this clear?

Questions to ask in the Polishing Phase

- Who is the intended audience for our research product(s)?
    - Do we have more than one audience?
- What is the next step in our research?
- Where do we plan to publish?
    - Can we turn our work into more than one publishable product?
- When do we expect a research product to be ready for a broader audience?
    - Consider products throughout the entire workflow.
- Why should we decide to build a software tool based on our work?
    - See suggestions in the Tool Development Guide (Box 4).

## Box 3 - Products

***Box 3 caption:*** *Research products can be developed in each of the three workflow phases. This box helps identify some options for each phase, including products less traditional to academia.*

Potential Products in the Exploration Phase

1. Publication of cleaned and processed data set (DOI)
2. Citable GitHub repository and/or computational notebook (e.g. data cleaning/processing, exploratory data analysis) (DOI)
3. GitHub Gists (e.g. particular piece of processing code)
4. White paper (e.g. explaining a dataset)
5. Blog post (e.g. detailing exploratory process)
6. Teaching/training materials (e.g. data wrangling)
7. Preprint (e.g. about a dataset or its creation) (DOI)
8. Peer-reviewed publication (e.g. about a curated dataset) (DOI)

Potential Products in the Refinement Phase

1. White paper (e.g explaining preliminary findings)
2. Literate Programming Notebooks (e.g. Jupyter Notebook, Knitr, Literate, etc.) (DOI)
3. Blog post (e.g. explaining findings informally)
4. Teaching/training materials (e.g. using your work as an example to teach a computational method)
5. Pre-print (e.g. preliminary paper before being submitted to a journal) (DOI)
6. Peer-reviewed publication (e.g. formal description of your findings) (DOI)
7. Grant application incorporating the data management procedure
8. Methodology (e.g. writing a methods paper) (DOI)

Potential Products in the Polishing Phase

1. Any and all of the potential products listed above for the Exploratory and Refinement Phases might serve as products in the Polishing Phase also (or instead!). (DOI applicable in many cases.)
    - Sometimes we reach the end of a given project before recognizing potential products. Revisit the list(s) of potential products that emerged from prior phases.
2. A software tool.
    - This might include a package, library, or interactive web application.
    - See Box 4 for further discussion of this potential research product.

## Box 4 - Tool Development Guide

***Box 4 caption:*** *Creating a new software tool as the polished product of a research workflow is non-trivial. This box furnishes a series of guiding questions to help researchers think through whether tool creation is appropriate to project goals, domain science needs, and team member skill sets.*

1. Should I make a tool?
    a. Does a tool in this space already exist that can be used to provide the functionality/answer the research question of interest?
    b. Does a new tool add to the body of scientific knowledge?
        i. Does it formalize our research question?
        ii. Does it extend/allow extension of investigative capabilities beyond the research question that our existing script was developed to ask?
    c. Does creating a tool advance our personal career goals or augment a desired/necessary skill set?
    d. Do we have the resources required to develop a new tool and will this be valued/help us reach our career goals?
        i. Time?

ii. Funding (if applicable)?
           iii. Domain expertise?
           iv. Programming expertise?
           v. Collaborative research partners with either time, funding, or relevant expertise?
2. What tool should we create?
    a. Should we build on an existing tool or make a new one?
    b. What is the scope of a new tool?
        i. What research area is it designed for?
        ii. Who is the envisioned end user? (e.g. scientist inside our domain, scientist outside our domain, policy maker, member of the public)
        iii. What is the goal of the end user? (e.g analysis of raw inputs, explanation of results, creation of inputs for the next step of a larger analysis)
3. How should we structure the tool? Consider:
    a. Language choice
        i. What are field norms?
        ii. Is it accessible (free, open source)?
    b. Data structures and types
        i. What is the likely form and type of data input to our tool?
        ii. What is the desired form and type of data output from our tool?
        iii. Are there pre-existing structures that are useful to emulate, or should we develop our own?
    c. Platform or framework
        i. Is there an existing package that provides basic structure or building block functionalities necessary or useful for our tool, such that we do not need to reinvent the wheel?